\documentclass{llncs}
\usepackage{llncsdoc}
\usepackage{graphicx}
\usepackage{times}
\usepackage{inconsolata}
\usepackage{float}

\usepackage{macros}

\newcommand{\sys}{Teechan\xspace}  
\newcommand{\txs}{tx/s\xspace} 

\newcommand{\hash}{$\mathit{Setup}_{\mathit{Hash}}$\xspace}

\newcommand{\setup}{{\emph{setup}}\xspace}
\newcommand{\refund}{{\emph{refund}}\xspace}
\newcommand{\settle}{{\emph{settlement}}\xspace}

\newcommand{\alice}{{\emph{Alice}}\xspace}
\newcommand{\bob}{{\emph{Bob}}\xspace}

\newcommand{\aenclave}{{$\mathit{TEE}_A$}\xspace}\newcommand{\benclave}{{$\mathit{TEE}_B$}\xspace}

\newcommand{\aid}{{$\mathit{ID}_A$}\xspace}
\newcommand{\bid}{{$\mathit{ID}_B$}\xspace}

\newcommand{\abtc}{{$\mathit{BTC}_A$}\xspace}
\newcommand{\bbtc}{{$\mathit{BTC}_B$}\xspace}

\newcommand{\atxo}{{$\mathit{UTXO}_A$}\xspace}
\newcommand{\btxo}{{$\mathit{UTXO}_B$}\xspace}

\newcommand{\apriv}{{$k_{\mathit{BTC}, A}$}\xspace}
\newcommand{\bpriv}{{$k_{\mathit{BTC}, B}$}\xspace}

\newcommand{\arsa}{{$\mathit{K}_A$}\xspace}
\newcommand{\brsa}{{$\mathit{K}_B$}\xspace}

\newcommand{\arsasig}{{$\mathit{Sigk}_A$}\xspace}
\newcommand{\brsasig}{{$\mathit{Sigk}_B$}\xspace}

\newcommand{\aone}{{$\mathit{A}_1$}\xspace}
\newcommand{\ax}{{$\mathit{A}_X$}\xspace}

\usepackage{enumitem}

\newcount\ppnum
\newcommand\ppnumber[1]{%
        \ppnum=#1\relax
        \ifnum\ppnum<0
                $-$%
                \ppnum=-\ppnum
        \fi
        \let\pptemp\empty
        \loop\ifnum\ppnum>999
                \count255=\ppnum
                \divide\ppnum by1000
                \count255=\numexpr \count255 - 1000*\ppnum \relax
                \edef\pptemp{,\ifnum\count255<100 0\ifnum\count255<10 0\fi\fi
                             \the\count255 \pptemp}%
        \repeat
        \the\ppnum
        \pptemp
}

\newcommand{\negspace}{\vspace{-0.5\baselineskip}}
\newcommand{\snegspace}{\vspace{-0.25\baselineskip}}

\renewenvironment{thebibliography}[1]{%
  \begin{oldthebibliography}{#1}%
    \negspace
     \setlength{\parskip}{0.2ex}%
     \setlength{\itemsep}{0.1ex}%
}%
{%
  \end{oldthebibliography}%
}

\usepackage{marginnote}

\begin{document}

\title{Teechan: Payment Channels Using Trusted Execution Environments}

\author{Joshua Lind\inst{1} \and Ittay Eyal\inst{2}
        \and Peter Pietzuch\inst{1} \and Emin G\"un Sirer\inst{2}} 
\institute{Imperial College London \and Cornell University and the Initiative for CryptoCurrencies and Contracts (IC3)}

\maketitle

\sloppy 

\begin{abstract}
  Blockchain protocols are inherently limited in transaction throughput and
  latency. Recent efforts to address performance and scale blockchains have
  focused on off-chain \emph{payment channels}.  While such channels can
  achieve low latency and high throughput, deploying them securely on top of
  the Bitcoin blockchain has been difficult, partly because building a secure
  implementation requires changes to the underlying protocol and the ecosystem.

  We present \emph{\sys}, a full-duplex payment channel framework that exploits
  \emph{trusted execution environments}.  \sys can be deployed securely on the
  existing Bitcoin blockchain without having to modify the protocol. It:
  (i)~achieves a higher transaction throughput and lower transaction latency
  than prior solutions; (ii)~enables unlimited full-duplex payments as long as
  the balance does not exceed the channel's credit; (iii) requires only a
  single message to be sent per payment in any direction; and (iv) places at
  most two transactions on the blockchain under any execution scenario.

  We have built and deployed the \sys framework using Intel SGX on the Bitcoin
  network.  Our experiments show that, not counting network latencies, \sys can
  achieve~\ppnumber{2480} transactions per second on a single channel, with
  sub-millisecond latencies.
\end{abstract}
%


\section{Introduction}

Bitcoin has grown significantly in popularity since its release. The ability to
transfer funds over a trustless, decentralized and global financial network has
attracted many industries and applications. As a result, adoption has grown rapidly, leading
to an exponential increase in the number of transactions sent per
day~\cite{transperday}. This growth exacerbates a natural problem: the Nakamoto
consensus protocol that underpins Bitcoin is fundamentally limited in
transaction throughput and imposes a significant minimum transaction
latency~\cite{croman2016scaling}. Furthermore, since miners must store the
history of every transaction ever made, accumulating storage costs increase the
cost of running nodes, which, in turn, leads to centralization pressure.

The maximum transaction throughput of Bitcoin is determined by the \emph{block
  size} and the \emph{block interval}. With a block size of 1~MB and an average
block interval of 10~minutes, Bitcoin can support a maximum of
7~\txs~\cite{wiki-scalability}. Recent proposals have suggested either tuning
these parameters, such as increasing the block size or reducing the block
interval~\cite{garzik2015bip100,andresen2015bip101,garzik2015bip102,wuille2015bipSipa};
or modifying the protocol, for example, by incrementally creating blocks so as
to avoid centralization bottlenecks and increase throughput~\cite{eyal2016ng}.
The former approach cannot scale Bitcoin by more than one order of magnitude, while
the latter requires changes to the underlying protocol, which practitioners
have been reticent to make. Other research suggests that hardware limits, such
as the cost of signature verification and storage latencies, cap Bitcoin to
200~\txs~\cite{decker2015fast}.

To handle demanding workloads, such as credit card processing ($\ge$ 10,000
\txs), recent proposals have focused on moving transactions off the blockchain
through the use of point-to-point \emph{payment
  channels}~\cite{bitcoin-contracts,decker2015fast,poon2016bitcoin}. Payment
channels allow for efficient, trustless fund transfers, in which parties can
exchange transactions without having to impact the blockchain except when the
channel is \emph{established} or \emph{terminated}. Consequently, two parties
can engage in a large number of fund transfers, only settling the net result on
the blockchain.  This decreases transaction confirmation latency, as only two
entities are involved, and reduces the load on the blockchain and network:
throughput scales linearly with the number of channels.

Despite the advantages of bidirectional payment channels, none have been
deployed securely on the existing Bitcoin network, as they assume modifications
to the underlying protocol. Specifically, they require transaction IDs to be
set before they are signed---a proposal to address this issue,
SegWit~\cite{segwit}, is currently mired in controversy~\cite{segwit-contro}.



We present \textbf{\sys}, the first \emph{high-performance micropayment
  protocol} that supports practical, secure, and efficient fund transfers on
the current Bitcoin network. Similar in design to Duplex Micropayment Channels
and the Lightning Network, \sys uses multi-signature time-locked transactions
to establish long-lived payment channels between two mutually distrusting
parties.  It fundamentally differs from existing protocols, however, in that it
leverages \emph{trusted execution environments}~(TEEs) to strengthen the
guarantees provided by the framework: (i)~\sys does not require any changes to
the Bitcoin network; (ii)~it enables infinite channel reuse as long as the
balance does not exceed the channel limits; and (iii)~it is time- and
space-efficient, requiring only one-way messages for sending payments and two
transactions to be placed on the blockchain in total. Section~\ref{sec:related}
discusses the comparison to prior art in detail.



At a high level, current implementations of TEEs can provide confidentiality
and integrity guarantees for code and data, but cannot guarantee liveness or
safe termination for a protocol. \sys is designed in a manner that, despite
these limitations, no party can gain access to more funds than their current
net balance. In particular, the TEE ensures that the private keys that control the
channel are never exposed to untrusted software or hardware, ruling out a large
class of potential attacks. These guarantees are robust in the presence of
compromised privileged software, such as the operating system, hypervisor, and
BIOS. In addition, an attacker who has full control of the hardware outside of
the CPU package, including the RAM, the system bus and the network, cannot
violate our security guarantees.

Overall, our paper makes the following contributions: (i)~it presents \sys, a
practical framework for low-latency, high-throughput, secure off-chain Bitcoin
transactions between mutually-distrusting parties;
(ii)~it describes the detailed operation of a prototype implementation of this
framework using Intel SGX as the TEE; and finally, (iii)~it presents
preliminary performance measurements from our prototype implementation,
demonstrating that \sys can achieve \ppnumber{2480}~\txs on a single payment
channel, thereby enabling system-wide aggregate throughput that can compete
with and surpass the requirements of credit card payment networks.



\section{Background}

In this section, we provide background on the technologies that underpin
\sys. We first give a short overview of Bitcoin, explore why it is unable to
scale, and then describe a trusted execution environment as provided by Intel
SGX.

\negspace
\subsubsection{Bitcoin}

Bitcoin~\cite{nakamoto2008bitcoin} is a distributed peer-to-peer network that
executes a replicated state machine. Each peer, or \emph{node}, in the network
maintains and updates a copy of the Bitcoin \emph{blockchain}, an append-only
log that contains the transaction history of every account in the network.
Users interact with the network by issuing \emph{transactions} to transfer
Bitcoins (\emph{BTC}). Valid transactions consume unspent transactions as
\emph{inputs} and create new unspent \emph{outputs} that can later be used in a
new transaction. To spend an \emph{unspent output}, a condition specified by a
\emph{locking script} must be met. Typically, a signature matching an address
proves that the user spending the output owns the account claiming the funds.
More complex locking scripts can be expressed, such as \emph{m-of-n multisig}
transactions, where \emph{m} signatures are required out of \emph{n} possible
signatures to spend the funds; and \emph{timelocked transactions}, which can
only be spent after a point in the future.

Transactions are appended to the Bitcoin ledger in batches known as
\emph{blocks}. Each block includes a unique ID, and the ID of the preceding
block, forming a chain. Peers in the network compete to generate and append
these blocks to the blockchain. This process, known as \emph{mining}, is
computationally expensive and requires solving a cryptographic puzzle.  Miners
are compensated for their efforts via the \emph{block reward} as well as the
\emph{transaction fees} collected from the transactions in that block.  The
Bitcoin protocol dynamically adjusts the difficulty of the cryptographic puzzle
so that a block is appended to the blockchain at an average rate of one block
every ten minutes.  In cases in which there are multiple blocks with the same
parent (\emph{forks}), the network adopts the chain with the greatest
difficulty.

This protocol architecture protects against \emph{double spend} attacks. In
such an attack, two conflicting transactions claim the same unspent outputs.
The Bitcoin protocol will ensure that the miners will mine at most one of these
transactions, and clients of the network will wait for additional succeeding
blocks (typically, 6) to guard against forks and
reorganizations~\cite{six-confirmations}.

Overall, the Bitcoin protocol suffers from two fundamental limitations.  First,
because it limits the size of each block and the rate of block generation, the
network is fundamentally limited in throughput.  Second, because the suffix of
the blockchain is subject to reorganization, users must wait until their
transactions are buried sufficiently deeply, incurring a minimum latency.

\negspace
\subsubsection{Trusted Execution Environments with Intel SGX}

Intel's \emph{Software Guard Extensions}~(SGX)\cite{sgx,sgx14,costanintel}
enable application code to be executed with confidentiality and integrity
guarantees. SGX provides \emph{trusted execution environments} known as secure
\emph{enclaves} that isolate code and data using hardware mechanisms in the
CPU. Assuming the physical CPU package is not breached, SGX enclaves are
protected from an attacker with physical access to the machine, including
access to the memory, the system bus, BIOS, and peripherals.

During execution, enclave code and data reside in a region of protected memory
called the \emph{enclave page cache}~(EPC).  When enclave code and data is
resident on-chip, it is guarded by CPU access controls; when it is flushed to
DRAM or disk, it is encrypted. A memory encryption engine encrypts and decrypts
cache lines in the EPC as they are written to and fetched from DRAM. Enclave
memory is also integrity-protected, ensuring that modifications and rollbacks
can be detected, and the enclave can terminate execution. Only code executing
inside the enclave is permitted to access the EPC. Enclave code can, however,
access all memory outside the enclave directly. As enclave code is always
executed in user mode, any interaction with the host OS through system calls,
e.g., for network or disk I/O, must execute outside the enclave.  Invocations
of the enclave code can only be performed through well-defined entrypoints
under the control of the application programmer.


In addition, SGX supports \emph{remote attestation}~\cite{johnson2016intel},
which enables an enclave to acquire a signed statement from the CPU that it is
executing a particular enclave with a given hash of memory, known as a
\emph{quote}. A third-party attestation service, e.g., as provided by the
\emph{Intel Attestation Service}~(IAS), can certify that these signed
statements originate from authentic CPUs conforming to the SGX specification.



\negspace
\section{Model and Goals}

Payment channels are applicable when two parties have long-lived financial
relationships that require frequent interaction with high-throughput, low
latency, and privacy guarantees. The central goal for \sys, then, is to
construct a duplex payment channel between two such endpoints, assuming that
these endpoints are equipped with trusted execution environments.

\negspace
\negspace
\subsubsection{Threat Model and Assumptions}

Our threat model assumes that both parties wish to exchange funds but mutually
distrust one another. Each party is potentially malicious, i.e., they may
attempt to steal funds, avoid making payments, and deviate from the agreement
if it benefits them. Any time during channel establishment, execution, and
closure, each party may drop, send, record, modify, and replay arbitrary
messages in the protocol. Either party may terminate the channel at any time,
and failures are possible.

We assume that each party has a TEE-capable machine and trusts the Bitcoin
blockchain, its own environment, the local and remote TEEs, and the code that
executes the \sys duplex channel protocol. The rest of the system, such as the
network between the parties and the other party's software stacks (outside the
TEE) and hardware are untrusted.  During protocol execution, any party may
therefore: (i)~access or modify any data in its memory or stored on disk;
(ii)~view or modify its application code; and (iii)~control any aspect of its
OS and other privileged software. 


Our threat model does not take into account denial-of-service attacks or
side-channel attacks. In practice, these are difficult to exploit, possible to
mitigate, and the subject of separate work outside the scope of this paper.

\negspace
\subsubsection{Goals}
            
A payment channel should operate as follows. A channel is established with a
\emph{setup transaction} in the blockchain to which each party deposits an
amount as credit.  While the channel is open, each party can pay its
counterparty via transaction messages sent from the payer to the payee.  A
payment can only be claimed if it was granted by a party, that is, theft should
not be possible.  At any point in time, the channel has a balance that must
reflect the difference between the amounts paid in each direction. The balance
should never exceed the credit in either direction. Either party can terminate
the channel at any time and settle the balance with a terminating transaction
that it places in the blockchain. The terminating transaction reflects a
balance that comprises all payments made by the terminator and all payments
received by the terminator from its counterparty.  Failures should only
negatively impact the party who failed.

Parties should only need to synchronize with the Bitcoin network during channel
establishment and at the point of settlement.  In particular, they should not
need to monitor the blockchain during the lifetime of the channel.

\section{\sys}

The intuition behind \sys is to exploit trusted execution environments~(TEEs)
to act as a trusted third party between two parties, \alice and \bob.

At a high-level, \sys works as follows. First, at setup, the TEE at each party
is securely given mutual secrets belonging to both parties. These secrets can
be used at any time to settle the channel, without needing cooperation. Next,
while the channel is open, the TEEs maintain channel state internally, free
from tampering due to the guarantees of trusted execution. Updates (payments)
are performed through a secure interface. Finally, \sys leverages secure
execution to settle the channel at termination.  Only on termination does a TEE
generate a Bitcoin transaction that can be placed in the blockchain.

Unlike previous approaches~\cite{decker2015fast,poon2016bitcoin}, \sys does not
make a settlement transaction available until channel termination. The
availability of such a transaction is the root cause behind much of the
complexity of today's payment channel implementations: it causes race
conditions, requires a timely response when leaked to the network prematurely,
and requires additional infrastructure for monitoring. This factoring of
crucial channel functionality into TEEs yields a simple and efficient
approach.

\begin{figure}[H]
  \centering
      \includegraphics[width=0.8\textwidth]{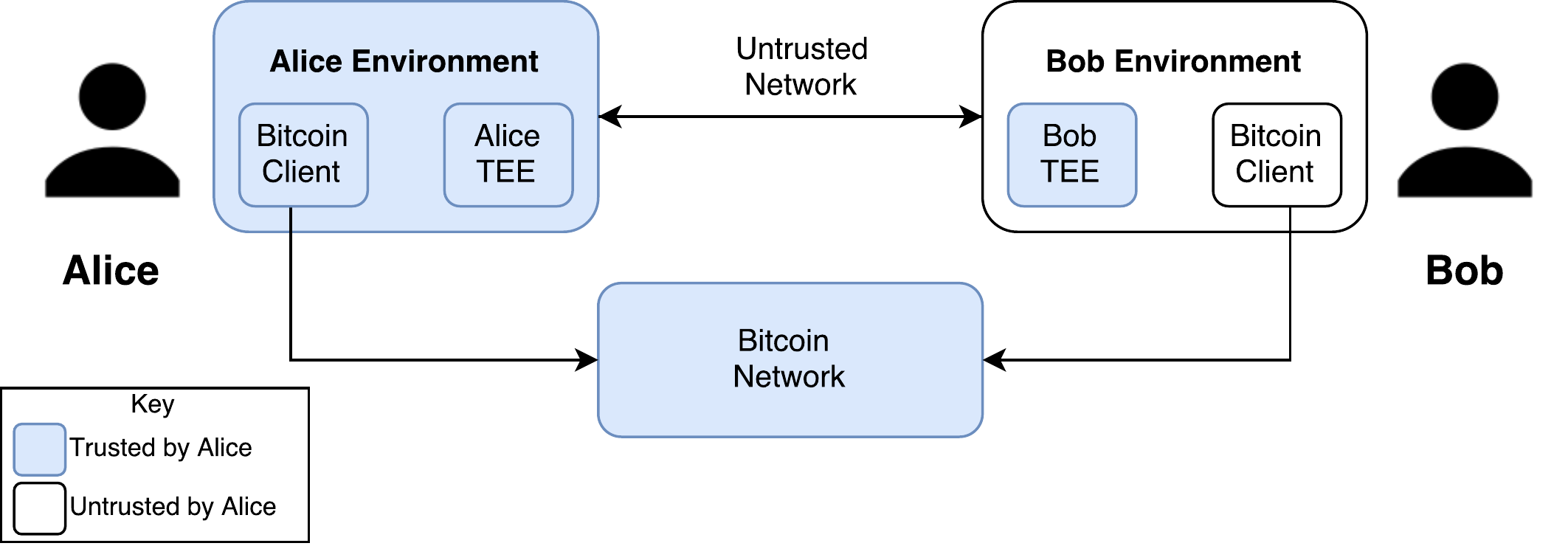}
  \caption{\sys Duplex Channel Architecture (Entities trusted by \alice are shaded.)} 
  \label{fig:arch}
\end{figure}

Figure~\ref{fig:arch} shows the \sys duplex channel architecture.  Both \alice
and \bob run their own TEEs alongside a connection to the Bitcoin network.
This connection is only used during channel establishment and closure.  The
figure highlights the entities trusted by \emph{Alice}.  An identical figure
can be constructed for \emph{Bob} using symmetry of the channel. We next
describe the protocol, and informally analyze its security in
Section~\ref{sec:security}.


\subsection{Protocol}

The \sys channel protocol operates in three phases: (i)~\emph{channel
  establishment}, (ii)~\emph{channel operation}, and (iii)~\emph{channel
  settlement}.  Figure~\ref{fig:protocol} shows the messages exchanged during
each of these phases in detail.  \emph{Alice}, \emph{Bob}, \emph{Alice's TEE}
(denoted \aenclave) and \emph{Bob's TEE} (denoted \benclave) are modeled as
separate entities.

For simplicity, we ignore mining fees in our example, although they are
supported in our implementation and affect only the initial setup and the final
settlement transactions.

\subsubsection*{A. Channel establishment} \vspace{-1em}

In the first phase, \sys establishes the duplex payment channel between \alice
and {\bob}.  Similar to prior
work~\cite{bitcoin-contracts,decker2015fast,poon2016bitcoin}, we construct a
payment channel using \setup and \refund transactions.  Both \alice and \bob
deposit funds into a \emph{2-of-2 multisig} Bitcoin address, forming a
\emph{setup} transaction.  A \emph{refund} transaction is constructed that
spends the \setup transaction and returns \alice and {\bob}'s deposits back to
them.  The \refund transaction is bounded by a lock-time using the
nLockTime~\cite{nlocktime} transaction field, making it valid only starting
some time in the future. The channel must be terminated prior to this time,
otherwise either party can terminate the channel as if no transactions took
place.

\begin{enumerate}[label=A\arabic*.]

\item First, \alice and \bob each provision their TEEs to construct \setup and
  \refund transactions. This requires: (i)~their Bitcoin \emph{private keys},
  \apriv and {\bpriv}; (ii)~the \emph{unspent transactions outputs sets} that
  they wish to include in the \setup transaction, \atxo and {\btxo}; and
  (iii)~the amount to deposit in the \setup transaction, \abtc and
  {\bbtc}.\label{itm:setup}

\item Second, \aenclave and \benclave establish a secure communication channel,
  authenticating each other through remote attestation.  To achieve this, each
  TEE generates an asymmetric encryption key pair and a random secret key using
  a secure random number generator. \aenclave binds the generated asymmetric
  public key to a quote, and sends it to Bob.  Using this quote, Bob can then
  verify that any message encrypted under \arsa can be decrypted solely by
  \aenclave, and that \aenclave is running the desired \sys code with the
  requisite binary hash.  The same is done in the reverse direction, so
  \aenclave obtains Bob's public key.  Upon successful mutual verification,
  \aenclave and \benclave know that any data encrypted under \arsa and \brsa
  can only be read by the opposite TEE. This establishes a confidential
  communication channel.

\item \benclave then presents its random secret key (denoted {\bid}), along
  with Bob's \setup data that it received in step~A1, to \aenclave. A signature
  over this message, under the private key of \benclave (denoted \brsasig), is
  also presented to ensure that it came from \benclave.  \aenclave generates
  the signed \setup and \refund transactions internally, and reveals to Alice
  the hash of the \setup transaction, denoted \hash, as well as the \refund
  transaction.  Only \aenclave knows the \setup transaction at this point.

  \aenclave then presents its random secret key {\aid}, along with Alice's
  \setup data that it received in step~A1 and the corresponding signature
  \arsasig, to \benclave. \benclave generates the \setup and \refund
  transactions internally, and reveals both to Bob.  Bob then broadcasts the
  \setup transaction onto the blockchain, establishing the channel.  Alice is
  notified of channel establishment by noting a transaction matching \hash on
  the blockchain. 

  Note that Bob could maul the \setup transaction before broadcasting it,
  making Alice's \refund transaction invalid. In this case, Alice presents the
  mauled \setup transaction to \aenclave to issue a new \refund transaction
  that closes the channel immediately. This requires that keys should never be
  re-used between separate channels, as is already a recommended good practice.
  In \sys, mauling the \setup transaction is equivalent to a denial-of-service
  attack.
\end{enumerate}

At the end of this three-step handshake, a secure communication channel is
established between the two TEEs. The slight asymmetry of the handshake is
critical for achieving the termination and loss properties described in
Section~\ref{sec:security}.

\subsubsection{B. Channel operation}\vspace{-1em}

\begin{figure}[H]
  \centering
      \includegraphics[width=1\textwidth]{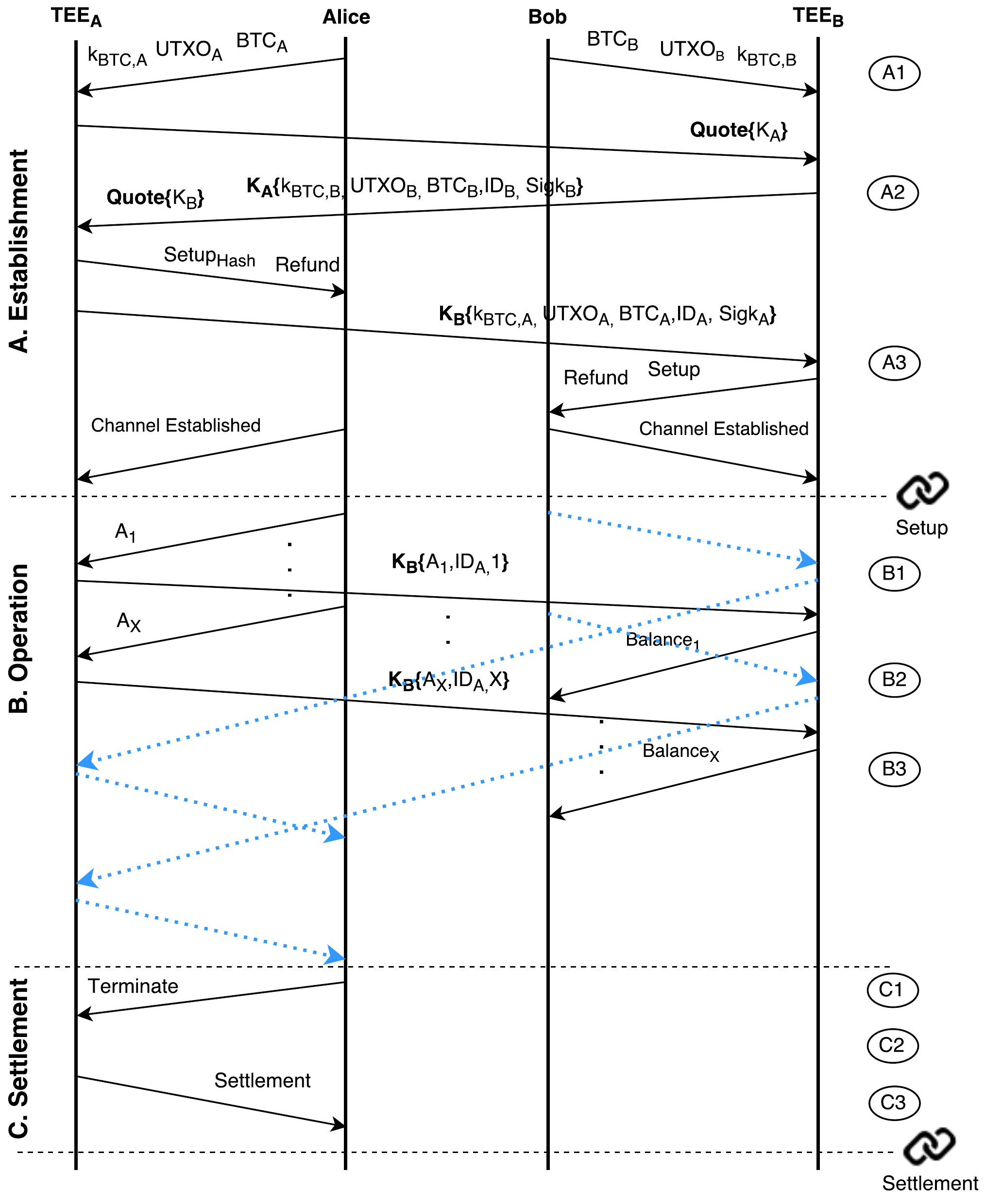}
  \caption{\sys Duplex Channel Protocol (For illustration purposes {\bob}'s payments to \alice are shaded, but unlabeled.)}
  \label{fig:protocol}
\end{figure}

Once a channel has been \emph{established} between \aenclave and \benclave,
\alice and \bob can begin exchanging funds.  In this phase, neither \alice nor
\bob need to maintain a connection with the Bitcoin network.  They can rapidly
make transactions through peer-to-peer updates. Note that in
Figure~\ref{fig:protocol}, payments made from \bob to \alice are shaded, but
unlabeled, for illustration purposes only. These payments exhibit the same
behavior, in a symmetric fashion, to the payments sent from \alice to \bob.

\begin{enumerate}[label=B\arabic*.]
\item To send funds to Bob, \alice sends a request locally to \aenclave,
  specifying the amount of Bitcoin that she wishes to transfer to Bob. These
  requests are denoted \aone through \ax, representing arbitrarily many payment
  requests.

\item When a TEE receives a payment request from the owner, it first checks
  that the current balance is greater than the amount to send.  If so, it
  updates the balance and generates a message authorizing the payment. The
  message contains the random secret key of the paying TEE \aid and the updated
  monotonic counter value.  The message is encrypted under the appropriate
  asymmetric public key \brsa.  Alice sends this message to Bob.

\item Bob receives the message and sends it to \benclave. Once the TEE receives
  the message, it decrypts it and asserts that it contains the correct secret
  key and that the value of the counter is greater by one than the previously
  presented counter. Then, it updates the balance and the counter for incoming
  messages. Finally, it notifies Bob of the new balance.

\end{enumerate}

Note that each party, outside the TEE, is in charge of maintaining a reliable
FIFO channel for the payment messages. This can be achieved with standard
go-back-n or similar protocols.  Tampering with the order of messages is
equivalent to a denial-of-service attack on the recipient only; the sender
always processes a payment. It is therefore in the best interest of the
receiver to ensure a reliable FIFO channel.

\subsubsection{C. Channel settlement}\vspace{-1em}

The final stage of the \sys protocol is \emph{channel settlement}. In this
phase, the payment channel is closed, and a valid transaction settling the
balance between \alice and \bob is broadcast to the Bitcoin network, thus
releasing the funds in the \setup transaction.

\begin{enumerate}[label=C\arabic*.]
\item At any point during phase~B, either party may send a \emph{terminate}
  request to their TEE.

\item Once a TEE receives a \emph{terminate} request from its owner, it
  generates a \settle transaction signed with \apriv and \bpriv, which spends
  the funds held in the \setup transaction according to the current channel
  balance.  It returns this transaction to the host, destroys all state held in
  TEE memory and halts its execution.

\item The party then forwards this to the Bitcoin network to complete the
  settlement.
\end{enumerate}

Similar to the Lightning Network~\cite{poon2016bitcoin}, \sys payment channels
do not suffer from channel exhaustion. \sys enables infinite channel reuse:
Alice and Bob can send funds back and forth until channel timeout.

Termination of a channel at the end of its lifetime is also similar to prior
work. When the \refund transaction becomes valid, either party can choose to
broadcast the \refund transaction, or to settle the current state of the
channel, as described above.  Whichever transaction is confirmed by the Bitcoin
network dictates the outcome of the channel.

Note that a unilateral channel termination by Bob cannot harm Alice: he will
not be able to receive further payments from Alice, but the closed channel will
accurately reflect all payments of which Alice is aware.  If Bob fails to
broadcast the termination transaction to the Bitcoin network, Alice can
independently close it from her side.

\sys is not a consensus protocol, nor is it designed to solve the
Byzantine-Generals Problem---Alice and Bob may not agree on the termination
state, but Alice's termination state is guaranteed to be acceptable by Bob, and
vice-versa.


\negspace
        \subsection{Security} 
        \label{sec:security}
\snegspace

In this section, we provide the intuition behind the security properties of the
protocol. We defer formal proofs of security to the full paper.

Any time during channel establishment, execution and closure, each party may
drop, send, record, modify, and replay arbitrary messages in the protocol. As
such, we informally evaluate and discuss the security of our protocol against
malicious and misbehaving parties. Note that any external adversary in the
system, such as an attacker who has compromised the network, has fewer
privileges than the counterparty in the channel, and so can be subsumed by a
malicious counterparty. Arguing security against the opposite party in a
channel is strong enough to protect against any external adversary.

During channel establishment, each TEE is provisioned with sensitive \setup
data from both parties.  This is always performed through a secure interface,
encrypted with a key internal to the TEE.  Communication with the counter
party's TEE is only performed after verifying that it is indeed a TEE executing
the \sys code.  Finally, no party can access the \setup transaction before the
other party has the \refund transaction.  Therefore, at the end of channel
establishment both parties have the refund transaction and only the TEEs have
both secrets.

During channel operation, once a party receives a payment, the sending party's
TEE has already registered this payment. Therefore, and due to the counter
encoded in each payment message, a party cannot revert a payment that it has
made when settling the channel. Early termination can only prevent a party from
receiving future payments, not harming the other party.




\paragraph{Intel SGX} 

We implement \sys on Intel SGX. Intel SGX provides secure TEEs offering both
execution integrity and confidentiality against an attacker on the same
machine, even one with physical access. These hardware guarantees, coupled with
\sys's architecture, enable the resulting system to be resilient against an
array of attacks. The tight integration of SGX with the CPU ensures that the
cost to launch an attack, or even gather enough know-how to attempt one, are
orders of magnitude higher than the value expected to be stored in payment
channels. Given the current market share of Intel CPUs, users already
implicitly trust a single hardware manufacturer with their secret keys. We
repeat, however, that nothing in the \sys protocol is Intel specific, and our
protocol can be ported easily to, for example, a Ledger hardware security
module~\cite{ledger}.
 
Replay attacks are detrimental to \sys security: if Alice could revert the
system to an old state, she could take a snapshot when the balance is in her
favor, and after sending payments to Bob, revert to that old state and settle
the channel at a wrong balance. SGX protects running enclaves against replay
attacks by protecting persistently stored snapshots from rollback attacks
through non-volatile hardware monotonic counters, which prevent a stale enclave
snapshot from being reused.

In our \sys prototype, if Alice fails, she can either ask Bob to settle at the
current balance, or wait until the refund transaction becomes available. It is
straightforward to extend the prototype such that the enclaves persist their
state to secondary storage, encrypted under a key and stored with a
non-replayable version number from the hardware monotonic counter. Our current
implementation, at the time of writing, does not leverage hardware monotonic
counters, because, while the counters are fully supported by the existing
hardware under Windows~\cite{windows-sgx-sdk}, the current SGX Linux
SDK~\cite{sgx-sdk} does not expose them yet. Porting our protocol to Windows or
support for monotonic counters in the Linux SDK can address this.


Currently, the validity of an Intel SGX attestation is certified through the
Intel Attestation Service~(IAS), which ensures that the quote originated from a
genuine SGX-capable Intel CPU. In our prototype, we do not use a trusted
connection between the enclave and the IAS; the quote is verified in untrusted
code, executed by the owner of the enclave during the setup phase.  This is
benign because misbehavior by a party at this stage would only harm that party,
as it would expose their private keys to a fraudulent remote enclave.
Terminating the TLS connection to the IAS inside the enclave~\cite{zhangtown}
would avoid this issue, but it is unnecessary under our trust model and would
needlessly increase the trusted computing base.


\snegspace
\section{Implementation \& Evaluation}
\negspace

We evaluate \sys using Intel SGX on the Bitcoin testnet. Our implementation is
fully compatible with the standard Bitcoin network.  We report preliminary
measurements from this implementation to illustrate the range of achievable
performance.

\negspace
\negspace
\subsubsection{\sys Implementation}

The \sys prototype has two components: a Bitcoin client and an Intel SGX
enclave application that executes the secure \sys protocol. Each party in the
payment channel maintains and executes their own client and enclave. For the
Bitcoin client, we fork the open-source
\texttt{libbitcoin-explorer}~\cite{libbitcoin-explorer}, a C++ Bitcoin library
that communicates with the Bitcoin network. \texttt{libbitcoin-explorer} relays
transactions and requests to a
\texttt{bitcoin-server}~\cite{libbitcoin-server}, a full Bitcoin peer in the
Bitcoin network. In our experiments, we use
\texttt{libbitcoin-explorer}~version 3.0.0 and communicate with a set of
public-facing \texttt{bitcoin-servers}~\cite{public-obelisk}.

For the \sys enclave application, we port a subset of \texttt{Bitcoin Core}
version~0.13.1~\cite{btc-core} to Intel SGX. Only some features of Bitcoin core
are needed inside the enclave: (i)~multisig address generation;
(ii)~transaction creation; (iii)~transaction signing; and (iv)~signature
verification. For asymmetric encryption between enclaves, we use RSA with
4096-bit keys.
Both \texttt{libbitcoin-explorer} and the \sys enclave communicate over
TCP using a lightweight message queuing library
(\texttt{ZeroMQ}~version~4.2.1~\cite{zmq}).

\negspace
\negspace
\subsubsection{Experimental Setup}
To evaluate \sys, we run all experiments on a single machine, which forms a
channel between two parties communicating through network sockets. We use an
SGX-enabled 4-core Intel Xeon~E3-1280 v5 at 3.7~GHz with 64~GB of RAM, and
Ubuntu~14.04 with Linux kernel~3.19. We deactivate hyper-threading, compile the
applications using GCC~5.4.0 with \texttt{-O2} optimizations and use the Intel
SGX SDK~1.7.

\negspace
\negspace
\subsubsection{Performance}

We measure the time taken by \sys to perform each of the three phases of the
protocol. To measure the throughput, we emulate an exchange between two parties
in which each party sends and receives payments sequentially in lock-step.
We measure the time for 10 million transactions to be exchanged. These
measurements yield an upper bound for our current implementation, as they
eliminate network bandwidth and latency. We defer a thorough evaluation of \sys
under varying network conditions, enclave topologies, and transaction patterns
to the full version of the paper.


%

Channel establishment and final settlement times are bounded by the time to
place the transactions in the blockchain. Once the channel is set up, we
measure an average latency of 0.40~ms and an average throughput of 2480~\txs.


For the purpose of demonstration, we provide a reference to a \sys payment
channel that was established, operated, and settled on the Bitcoin test
network. Each side deposited 50~bitcoin in the \setup
transaction\footnote{http://tbtc.blockr.io/tx/info/d55dcfebff45d7e4f9970edd053c87cb0b659e459f4f6360d4a2c17837e79410},
and the channel was closed with a balance of 9~bitcoin for
Bob\footnote{http://tbtc.blockr.io/tx/info/1a736822a4f518eb137658030f1e11a804d64d1da48c195222f604aaf2df908e}.
A fee of 0.002~bitcoin was paid on each transaction.

%

\negspace
    \section{Related Work} \label{sec:related}
\negspace

Direct payments were proposed by Chaum~\cite{chaum1982blind} in ecash to achieve privacy. The ecash assumptions are significantly weaker than those offered by payment channels. Mainly, cheating is enforced in retrospect, through external punishment mechanisms. 

Several proposals address the performance issues of the Bitcoin network, from the GHOST protocol and alternatives to the chain structure~\cite{sompolinsky2015ghost,lewenberg2015inclusive,ethereum2015white}, to alternative block generation techniques~\cite{eyal2016ng,kogias2016byzcoin,pass2016hybrid}. 
Others~\cite{mazieres2015stellar,cachin2016architecture,miller2016honeybadger} build on classical consensus protocols~\cite{castro1999practical} or operate in permissioned settings. 
While they all improve on the Nakamoto blockchain performance, none can reach the performance offered by direct channels that do not require global system consensus for each transaction. 

Unidirectional Bitcoin \emph{micropayment channels} were first informally discussed by Hearn and Spilman~\cite{bitcoin-contracts}. 
These could not be deployed directly as they require changes to the Bitcoin protocol, unlike \sys.
Alternative proposals for unidirectional micropayment channels have been made to avoid these changes, however, all unidirectional payment channels only operate in a single direction and suffer from channel exhaustion. 

Decker and Wattenhofer~\cite{decker2015duplex} were the first to realize Duplex Micropayment Channels (DMC), improving the exhaustion limit. 
In DMC, two parties form a pair of channels, one in each direction, and re-balance them as needed, that is, when the credit in one direction is depleted but after there have been transactions in the opposite direction. 
However, the number of resets possible is limited at channel construction, depending on the time allotted for the refund timeout and the bound on the time to place a transaction on the blockchain. 
Therefore the total amount that can be sent on the channel in one direction is bounded by the deposit amount times the maximal number of resets. 
DMC also requires changes to Bitcoin. 

Additionally, on disagreement, $1 + d + 2$ transactions have to be placed on the blockchain,
where $d$ represents the invalidation tree's active branch. In \sys there is no limit on the total amount moving in any direction, and only 2 transactions are ever placed in the blockchain.

Lightning Network (LN)~\cite{poon2016bitcoin} allows for unlimited reuse of its channels. Two parties form a series of transaction structures, where each update invalidates the previous one. 
If a party tries to settle the channel on the blockchain with an invalidated state, its counterpart sees this transaction on the blockchain and can redirect all the deposited amount to itself. 
The performance impact of this protocol is that payments happen in a serial fashion, one at a time. 
Updating the balance takes about four message exchanges (from deciding on the new value to sending transaction signatures in a certain order). 
During these exchanges, no payments can be reliably made. 
In \sys a payment is done with a single message, and payments in both directions can be made concurrently, making it full-duplex rather than half-duplex. Additionally, on disagreement, The Lightning Network places four transactions in the blockchain, while \sys places only two.

Informal proposals have been sketched to potentially deploy Lightning Network on the Bitcoin network without any changes to the Bitcoin protocol. However, these come with various limitations; primarily, a channel can only be funded by a single party, and parties need to monitor the blockchain in order to react to invalidated states. This is not the case for \sys however, as both parties can deposit into a \sys
channel, and neither party ever controls a transaction that reflects an old state.

Lightning Network effort aims to construct a multi-hop network of payment channels, a topic that is outside the scope of this report. We believe the LN network structure can be made to work with \sys channels, a challenge we defer to future work.

Towncrier~\cite{zhangtown} uses a TEE to provide authenticated data feeds for smart contracts. 

\negspace
\negspace
    \section{Conclusion} 
\negspace    
    

We presented \sys, full-duplex payment channels based on the existing Bitcoin network with trusted execution environments. 
The \sys prototype, built on Intel SGX, can achieve \ppnumber{2480}~\txs and a transaction latency of 0.40ms in optimal conditions. 
It advances the state of the art by obviating the need to modify the underlying Bitcoin protocol for a practical deployment, improving channel performance, and reducing blockchain overhead.

\negspace
\subsection*{Acknowledgements}
This project received funding from the European Union’s
Horizon 2020 research and innovation programme under the SecureCloud
(Grant agreement No. 690111) project.

\clearpage
\bibliographystyle{acm} 
\bibliography{references,btc} 
\end{document}